\documentclass[runningheads]{llncs}
\usepackage{cite}
\usepackage{amsmath,amssymb,amsfonts}
\usepackage{graphicx}
\usepackage{textcomp}
\usepackage{algorithm}
\usepackage{cite}
\usepackage{siunitx}
\usepackage{amsmath,amssymb,amsfonts}
\usepackage{algorithm}
\usepackage{graphicx}
\usepackage{textcomp}
\usepackage{xcolor}
\usepackage{moreverb}
\usepackage{tabularx}

\usepackage{threeparttable} 
\usepackage{multirow} 
\usepackage{rotating} 
\usepackage{bigstrut} 
\usepackage{todonotes}
\usepackage[caption=false]{subfig}
\usepackage{graphicx}
\usepackage[utf8]{inputenc}
\usepackage{paralist}
\usepackage{lipsum}
\usepackage{multirow}
\usepackage[normalem]{ulem}
\usepackage{arevmath}     
\usepackage[noend]{algpseudocode}
\usepackage[T1]{fontenc}
\usepackage{textcomp}
\usepackage{xcolor}
\usepackage{soul}

\usepackage{todonotes}
\newboolean{showcomments}
\setboolean{showcomments}{true}         

\ifthenelse{\boolean{showcomments}}
{\newcommand{\nb}[2]{
		\fbox{\bfseries\sffamily\scriptsize#1}
		{\sf\small$\blacktriangleright$\textit{\textcolor{red}{#2}}$\blacktriangleleft$}
	}
}
{\newcommand{\nb}[2]{}
	
}

\def\BibTeX{{\rm B\kern-.05em{\sc i\kern-.025em b}\kern-.08em
    T\kern-.1667em\lower.7ex\hbox{E}\kern-.125emX}}
\begin{document}

\title{Mitigating and Analysis of \\Memory Usage Attack in IoE system
}

\author{Anonymous}
\author{Zainab Alwaisi\inst{1}
\and
Simone Soderi\inst{1,2}
\and
Rocco De Nicola\inst{1,2}
}
\authorrunning{Z. Alwaisi et al.}

\institute{IMT School for Advanced Studies, Lucca, Italy\\
\email{\{zainab.alwaisi, simone.soderi, rocco.denicola\}@imtlucca.it} \and
CINI Cybersecurity Laboratory, Roma, Italy 
}

\maketitle

\begin{abstract}
Internet of Everything (IoE) is a newly emerging
trend, especially in homes. Marketing forces toward smart homes are also accelerating the spread of IoE devices in households. An obvious risk of the rapid adoption of these smart devices is that many lack controls for protecting the privacy and security of end users from attacks designed to disrupt lives and incur financial losses. Today the smart home is a system for managing the basic life support processes of both small systems, e.g., commercial, office premises, apartments, cottages, and largely automated complexes, e.g., commercial and industrial complexes. One of the critical tasks to be solved by the concept of a modern smart home is the problem of preventing the usage of IoE resources. Recently, there has been a rapid increase in attacks on consumer IoE devices. 

Memory corruption vulnerabilities constitute a significant class of vulnerabilities in software security through which attackers can gain control of an entire system. Numerous memory corruption vulnerabilities have been found in IoE firmware already deployed in the consumer market. This paper aims to analyze and explain the
resource usage attack and create a low-cost simulation environment to aid in the dynamic analysis of the attack. Further, we perform controlled resource usage attacks while measuring resource consumption on resource-constrained victims' IoE devices, such as CPU and memory utilization. We also build a lightweight algorithm to detect memory usage attacks in the IoE environment. The result shows high efficiency in detecting and mitigating memory usage attacks by detecting when the intruder starts and stops the attack.


\end{abstract}

\newpage
\noindent\rule{8.4cm}{1pt}\\
Kindly reference this version of the paper:\\

Alwaisi, Z., Soderi, S., De Nicola, R. (2023). Mitigating and Analysis of Memory Usage Attack in IoE System. In: Vo, NS., Tran, HA. (eds) Industrial Networks and Intelligent Systems. INISCOM 2023. Lecture Notes of the Institute for Computer Sciences, Social Informatics and Telecommunications Engineering, vol 531. Springer, Cham. https://doi.org/10.1007/978-3-031-47359-3\_22
\\
You can reference the following BibTeX entry:
\begin{verbatim}
@InProceedings{10.1007/978-3-031-47359-3_22,
author="Alwaisi, Zainab
and Soderi, Simone
and De Nicola, Rocco",
editor="Vo, Nguyen-Son
and Tran, Hoai-An",
title="Mitigating and Analysis of Memory Usage Attack in IoE System",
booktitle="Industrial Networks and Intelligent Systems",
year="2023",
publisher="Springer Nature Switzerland",
address="Cham",
pages="296--314",
isbn="978-3-031-47359-3",
doi="10.1007/978-3-031-47359-3_22".
}



\end{verbatim}
\noindent\rule{8.4cm}{1pt}

\section{Introduction}
\begin{figure}[!ht]
	\includegraphics[width=9cm, height=7cm]{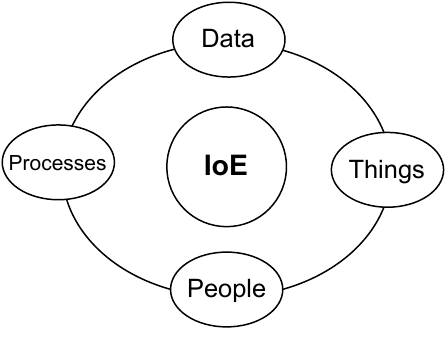}
	\centering
	\caption{The Definition of Internet of Everything (IoE).}
	\label{FIG:IoE}
\end{figure}

The Internet of Everything (IoE) encompasses data, people, the Internet of Things (IoT), and processes. IoE builds on IoT, which focuses on connecting network devices equipped with specialized sensors or actuators through the Internet~\cite{jamil2022}. The sensors and actuators can detect and respond to environmental changes, including light, temperature, sound, vibration, etc. IoE dramatically expands the scope of IoT by adding components that can provide richer experiences for businesses, individuals, and countries. For example, instead of simply relying on things to interact with their environments, as shown in Figure~\ref{FIG:IoE}, IoE can leverage all related data and processes to make IoT more relevant and valuable to people~\cite{zhan2022ioe}. The ultimate goal of IoE is to boost operational efficiency, offer new business opportunities, and improve the quality of our lives. Better to relate to this idea; take the scenario of a person uncertain about closing a gas valve at home. An IoE solution allows a user to automatically check the gas valve's status and close it remotely if necessary~\cite{Ryoo2017}~\cite{shi2022intelligent}.

Despite its potential rewards, IoE could pose significant security threats to its adopters. The number of IoE devices around us is steadily increasing, and IoE is starting to play a more critical role in our everyday lives. In particular, the link between the physical world and cyberspace established by IoE increases the risk of cyber attacks targeting smart devices since attacks against IoE can directly impact the health~\cite{rani2021towards} and the welfare of their end users. Building on our gas value scenario, one can easily imagine a threat scenario in which an attacker causes a gas leak on purpose~\cite{zhang2022near}.
Even more alarming is that we are often oblivious to the quantity and nature of the IoE devices surrounding us, not to mention the potential security risks they represent. The recent security incidents resulting from IoE security vulnerabilities corroborate this observation. In particular, one of them is a Distributed Denial of Service (DDoS) attack against Dyn~\cite{DDoSOnline16} in October 2016. This incident involved a botnet called Mirai, consisting of approximately $100,000$ IoE hosts, including digital cameras and routers. The Mirai botnet launched DDoS attacks against Dyn, bringing down its Domain Name Servers (DNS), which resulted in an outage of major commercial websites, e.g., Netflix and CNN~\cite{kathole2022energy}.
Another attack could affect the sensors and actuators' resources, making the smart devices unavailable to end-users.

Due to these emerging threats, it is imperative to raise awareness of potential IoE security risks~\cite{wei2019convergence} among end users through systematic risk assessment and effective visualizations. Home users are especially vulnerable because they are increasingly surrounded by IoE appliances, e.g., hands-free speakers, baby monitors, and security cameras. However, they lack the resources and skills to identify their IoE-related threats, remediate them, and minimize potential security risks. Therefore, in this paper, we mainly focus on analyzing the memory usage attack in smart devices and mitigating the effect of this attack by building a lightweight algorithm to detect memory usage attacks by calculating the memory usage of the smart device.

To accomplish this goal in home networks, we first identify
memory usage attacks in smart home devices. Next, we analyze the effect of the attack by sending malicious attacks to affect the resources of the smart devices and calculate its effect on memory usage. We then elicit and document threats in the form of threat scenarios. Once specified, we build a lightweight algorithm to detect and mitigate the effect of the attack on memory usage.

\subsection{Motivation and Contribution}    
\label{SUBSEC:MOTIVATION}

IoE is a fast-growing field with capabilities to revolutionize the whole industry. As per market trends, more than $20$ billion smart devices will be deployed in the next five years~\cite{Murtuza2022}. These interconnected devices will be generating sensitive data which needs to be protected. The field of IoE is making leaps and bounds technologically. There are multiple limitations while deploying IoE devices daily, e.g., battery life and lightweight computation. Therefore, building a novel
security mechanism aims to protect the functionalities and privacy of sensitive IoE network environments, including healthcare, smart cities, etc. However, due to
the substantial number of nodes in the environment and their
restricted computing capabilities, securing smart nodes in the IoE environment is essential to protect the data and make the devices available to end-users. A lightweight mitigation technique should be considered to protect smart devices from resource-constraint attacks such as Denial of Services (DoS), Distributed Denial of Services (DDoS), and other malicious attacks. Our main contribution is building a lightweight technique to detect memory usage attacks in smart devices deployed directly at sensors. It applies real-time memory usage calculation to discriminate between different memory usage, e.g., read/write to memory. In this work, we consider different behaviors on the memory of smart devices. We measure the memory usage when there is read and write, under or without the attack, to evaluate the best detection of memory usage attack. We simulate the mitigation technique and assess the results by applying the proposed technique to smart devices, such as the Raspberry Pi\footnote{https://www.raspberrypi.com/documentation/} and Arduino. We measure the current memory usage of the smart device to monitor the memory usage to discriminate between normal and abnormal behaviors. Therefore, this algorithm design is a protection strategy for IoE devices to maintain their integrity, seamlessly make them available to legitimate users, and protect them from memory attacks by considering their resource constraints.

\subsection{Organization of the paper}    
\label{SUBSEC:Orgnazation}
We organized this paper as follows. Section~\ref{SEC:RW} presents a related work and background reading of resource-constrained attacks in IoT systems, e.g., memory usage attacks. We discuss the threat scenario and its analysis in Sections~\ref{SEC:TS} and~\ref{SEC:SA}. We describe our proposal, including metrics definition, methodology, and the detection algorithm, in Section~\ref{SEC:PA}. In Section~\ref{SEC:EXP}, we show the results and discussions. Finally, Section~\ref{ConFU} presents some concluding remarks and future works.
\begin{table*}[h]
\renewcommand{\arraystretch}{1.7}
	\centering
        \small
	\caption{Memory Usage analysis before and after the attack. 
 }
	\label{tab:memoryscan}
\begin{tabular}{|m{0.25\textwidth}<{\centering}|m{0.20\textwidth}<{\centering}|m{0.25\textwidth}<{\centering}|m{0.25\textwidth}<{\centering}|m{0.25\textwidth}<{\centering}}
\cline{1-4}
\multicolumn{1}{|c|}{\textbf{Device}}  & \textbf{Status}       & \textbf{\% CPU Usage} & \textbf{\% Memory Usage} &  \\ \cline{1-4}
\multirow{3}{*}{\textbf{Raspberry Pi}} & Idle         & \multicolumn{1}{c|}{$0.55\div0.88$}            & \multicolumn{1}{c|}{$10\div20$}                                              &  \\ \cline{2-4}
                                       & Active      & \multicolumn{1}{c|}{$0.88\div1.50$}               & \multicolumn{1}{c|}{$20\div35$}                                               &  \\ \cline{2-4}
                                       & Under Attack & \multicolumn{1}{c|}{$1.5\div16.5$}            & \multicolumn{1}{c|}{$36\div66$}                                          & 
                                       \\ \cline{1-4}

\multicolumn{1}{|c|}{\textbf{Device}}  & \textbf{Status}       & \textbf{Thread Time [s]} & \textbf{\% Memory Usage} &  \\ \cline{1-4}                                      
\multirow{3}{*}{\textbf{Arduino}} & Idle        & \multicolumn{1}{c|}{$1\div20$}             & \multicolumn{1}{c|}{$8\div11$}                                              &  \\ \cline{2-4}
                                       & Active      & \multicolumn{1}{c|}{$21\div45$}              & \multicolumn{1}{c|}{$11\div16$}                                              &  \\ \cline{2-4}
                                       & Under Attack & \multicolumn{1}{c|}{$\geq45$}           & \multicolumn{1}{c|}{$17\div45$}                                        
                                       
                                       &\\ \cline{1-4}
\end{tabular}
\end{table*}
\section{Related Work}
\label{SEC:RW}
The IoE links people, data, things, and processes to make interconnections easier and more far-reaching than ever before~\cite{Jara2013}. As such, everyday appliances should be subjected to rigorous cyber security testing to the same degree that these appliances are tested and measured for traditional qualities, e.g., durability, fit-for-purpose, maintenance, etc. Unfortunately, standardized and independent verification of IoE devices is in its nascent stage, with IoE security being the focus of legislation and standard security criteria.
Different authors tried to mitigate and detect attacks by analyzing the memory. 
Memory analysis has attracted several malware researchers. Vömel \emph{et al.} in~\cite{vomel2011survey} surveyed the main memory acquisition and analysis techniques. In~\cite{rathnayaka2017efficient}, Rathnayaka, et al. have observed that successful malware infection leaves a memory footprint. Zaki \emph{et al.} in~\cite{zaki2014unveiling} studied the artifacts left by rootkits at the kernel level, such as driver, module, System Service Dispatch Table (SSDT) hook, Interrupt Descriptor Table (IDT) hook, and callback. The experiments proved that certain activities, such as callback functions, modified drivers, and attached devices, are the most suspicious activities at the kernel level. In~\cite{aghaeikheirabady2014new}, Aghaeikheirabady presented an analysis approach that extracts features available in memory, such as function calls, Dynamic-Link Libraries (DLLs), and registry, and compares the information available in different memory structures to increase the accuracy. The approach relies on the frequencies of the extracted features to classify them, and an overall accuracy of $98$\% is measured by applying Naïve Bayes. However, a significant drawback is the high False Positive Rate (FPR) that exceeded $16$\%. 

Similarly, in~\cite{mosli2016automated}, Mosli \emph{et al.} introduced a technique that detects malware based on extracting three features from memory images; API calls, registry, and imported libraries. However, the experiments were performed on each feature individually, and maximum accuracy of $96$\% was achieved using the
Support Vector Machines (SVM) classifier on the registry activities feature. Afterward, in their following work~\cite{mosli2017behavior}, Mosli \emph{et al.} utilized the process handles available in memory to detect malware. The experiment has found that the handles used by malware are process handles, mutants, and section handles. However, when applying the random forest classifier,
their approach achieved a modest accuracy slightly higher than $91$\%. Likewise, Duan \emph{et al.}~\cite{duan2015detective} presented an approach to extract live
DLL is featured from memory and employed to detect malware variants that use the same DLLs. The experimental result showed an accuracy of $90$\% achieved using the hidden naïve
bayes classifier.
Furthermore, Dai \emph{et al.}~\cite{dai2018malware} proposed a malware detection and classification approach based on extracting memory images and converting them into fixed-size greyscale images. The approach
then extracted the features from the images, using a gradient histogram, and used them to classify malware. An accuracy of $95.2$\% was obtained using the neural network classifier. Moreover, the authors of this work previously combined API calls from behavior analysis and
memory analysis into one vector to represent each sample. A dataset was used, which consisted of $1200$ malware and $400$ benign files, to train the SVM classifier. The work confirmed that memory analysis could overcome the limitations of behavior analysis~\cite{sihwail2019malware}.

In this paper, we monitor the memory usage of the smart devices in the IoE environment to detect memory usage attacks and mitigate resource-constraints attacks. We also perform a testbed environment to measure the memory usage of the smart devices before and after attacking the IoE environment.
In the experiments, the effectiveness of the proposed approach on memory usage attack detection and classification has been demonstrated and measured by three evaluations: classification accuracy,
monitoring the memory usage, and detecting the attack.
To the best of our knowledge, this is the first work to examine and detect such kind of memory usage attack in IoE systems.

\section{Testbed Scenario}
We used Raspberry Pi and Arduino as smart home devices in this experiment. We used different software tools for attacking data generation and collection. On the adversary side, we used \emph{Nmap}\footnote{https://nmap.org/} to launch a network scan and identify devices' status, such as online or offline, IP address, and MAC address. Different tools can generate malicious attacks on the victim side, such as \emph{hping3}\footnote{https://www.kali.org/tools/hping3/}.
We used \emph{tshark} tool \footnote{https://www.wireshark.org/} to evaluate the impact of memory usage attacks on smart devices and capture WiFi traffic.

We also created a module inside the smart device to monitor memory usage and register all memory behaviors in the database (DB). The monitoring mode registers the behavior of the smart devices once it is \emph{Idle}, \emph{active}, and \emph{under attack}.
\begin{figure}[!ht]
	\includegraphics[width=10cm, height=10cm]{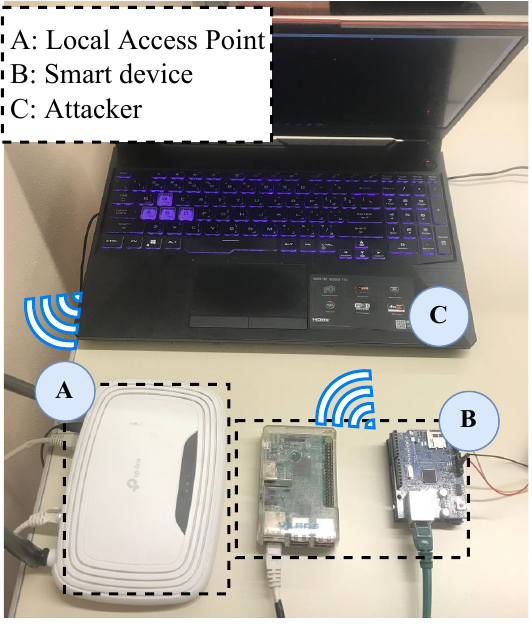}
	\centering
	\caption{Testbed scenario showing the devices used in our experiment.}
	\label{FIG:MUB1}
\end{figure}
Different stages are used to run our experiment. In the first stage, we monitor the memory usage
once the device is Idle, Active, and under attack. 
Then, we run a network scan to capture the port and device status. Once we ensure that the device is connected to the Internet, we send memory usage attacks for two purposes: first, to affect the memory, and second, to consume more memory usage and study the behavior of the attack. Then, we run memory usage monitoring to calculate the memory usage of the devices and study the devices' behaviors before and after the attack. 
\begin{figure}[ht]
	\includegraphics[width=\linewidth, height=8cm]{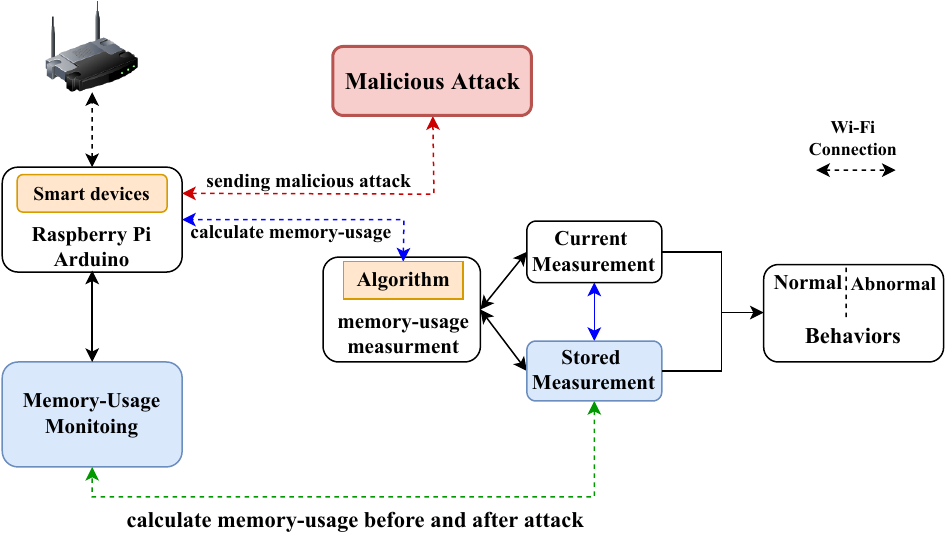}
	\centering
	\caption{Testing Environment.}
	\label{FIG:T2D}
\end{figure}

\section{Threat Scenarios and Threat Model}
\label{SEC:TS}
This section will briefly discuss the design of memory usage attacks on IoE smart devices. The memory usage attack aims to affect the smart device's memory by sending malicious attacks such as DoS or DDoS attacks. In particular, memory usage attack targets a specific type of vulnerable IoE and embedded devices because these devices have very little build-in-security protection and suffer from resource-constraints problems.

\subsection{Threat Scenario}
The smart devices of IoE suffer from low computation problems such as low energy and memory. The resource-constraints problems encourage attackers to attack these devices by flooding the smart devices with malicious attacks. In this paper, we assume that the attacker has gained access to the control network and can communicate with the smart devices as an insider threat (e.g., a consumer who uses current or past authorized access to the smart devices to exceed or misuse) or an external hacker. A wide range of attacks, such as malicious attacks, will be available for the attacker when he/she gains access to a control network. In this paper, we study memory usage attacks on the smart devices of IoE systems.

The threat scenario used in our experiment was first to scan the network and get different information about the port and devices' status. For scanning the network, we install \emph{Nmap} on Kali-Linux. In this scenario, the attacker can send a malicious attack to the smart device to affect its resources in terms of memory. The IP address, port, and device status are stored in the DB for further calculation. After scanning the network, we start the monitoring mode of the smart device's memory usage; once the device is Idle and active, and when we send a malicious attack using \emph{hping3} tools to the smart device. In this case, we study the memory behavior before and after attacking the smart devices. We also store all information about the memory, such as memory in total, memory usage, and CPU usage before and after attacking the smart devices.

In particular, the source code consists of three different parts:
\begin{itemize}
    \item Memory usage attack: This module commences with the DoS and DDoS attack to send malicious attacks to the smart devices and affect their memory.
    \item Scanner: This module scans the network and gets different information about the smart devices. Also, it sends the IP address of the attacked smart devices for further calculation
    \item Memory-monitoring-mode: This module monitors the memory usage of the smart devices when it is Idle, active, and under attack. The monitoring mode helps to register different memory behaviors for detecting such attacks.

\end{itemize}

\subsection{Threat Model}
We present a model of attacks on the memory usage of the smart devices of IoE systems, which can be used to understand the possible attack vectors intuitively and concisely. Also, to build lightweight algorithm~\footnote{https://github.com/developerZA/MitigationMemoryAttack.git} for detecting smart devices from these attacks.
Let us assume $ATK$ denotes the attacker while $d$ denotes the smart devices, and $MEM$ denotes the smart devices' memory usage. 
According to our model, every attack originates from an attacker $ATK$ where $atk\in ATK$ by a means towards a target $d$. We can model this relationship as follows:

\begin{equation} \label{eq1}
 ATK \mapsto mem \to D
\end{equation} 
where $atk \in ATK, d \subset D$, $mem \in MEM$. The notation $\mapsto$ maps the attacker ($ATK$) to the victim's ($D$) memory ($mem$). 
\begin{figure*}[ht]
	\includegraphics[width=\textwidth, height=9cm]{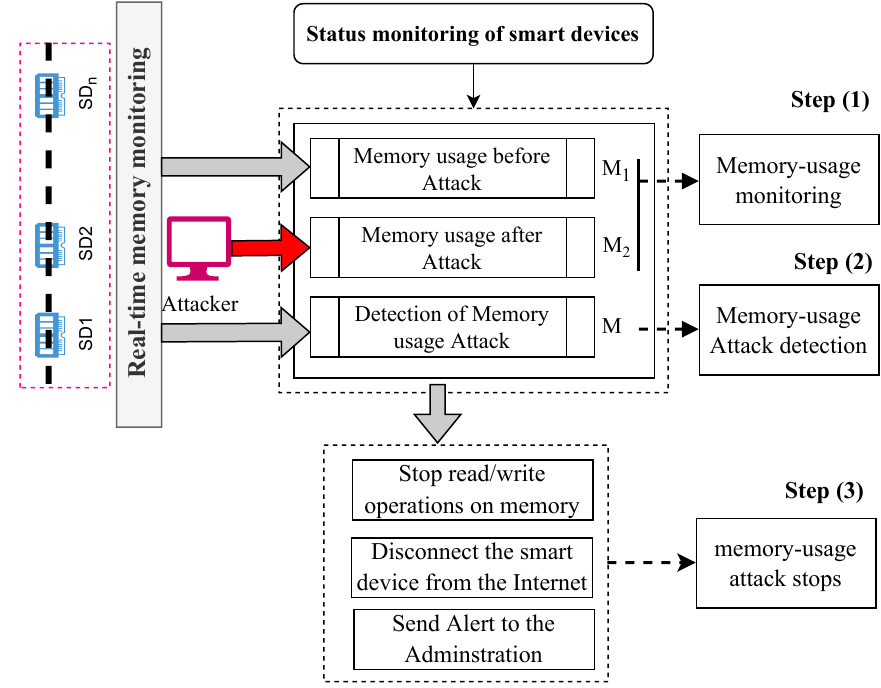}
	\centering
	\caption{Schematic diagram of the proposed method.}
	\label{FIG:F2D}
\end{figure*}
For calculating the memory usage and CPU usage of the smart devices before and after attacking the memory, the following equation math represents this calculation.
Let us describe the memory usage measurement  ($MEM$) footprints considering the set of different device statuses in the attack's absence or presence.
\begin{equation} \label{eq1}
 MEM(d) = f(mem(d),ATK, n) \quad \textrm{and} \quad n \in [0,1]
\end{equation} 

Where ($mem_{d}$) the memory usage measurement ($mem$) of the smart device ($d$) at a point in time in
the absence or presence of cyberattacks for a specific attack ($ATK$), and $n$ is the number of memory usage measurements in a time interval, $f(mem(d),ATK,n)\in[0,1]$ where $0$ is the minimum memory usage measurement, and $1$ presents the maximum memory usage measurement in the absence or presence of the attack.
The CPU ($CPU$) usage measurement is also calculated for the Raspberry Pi device as follows:

\begin{equation} \label{eq2}
 CPU(d) = f(cpu(d),ATK, n) \quad \textrm{and} \quad n \in [0,1]
\end{equation} 

Where ($cpu_{d}$) is the CPU usage measurement ($cpu$) of the smart device ($d$) at a point in time in
the absence or presence of cyberattacks for a specific attack ($ATK$), and $n$ is the number of CPU usage measurements in a time interval, $f(cpu(d),ATK,n)\in[0,1]$ where $0$ is the minimum CPU usage measurement, and $1$ presents the maximum CPU usage measurement in the absence or presence of the attack.

We do not calculate the CPU usage for the Arduino, as it is a microcontroller. We focus only on the maximum memory usage through or without the attack using a particular library called \emph{MemoryFree} and \emph{pgmStrToRAM}. And we also calculate \emph{micros()} or \emph{millis()} before and after sending the malicious attack. We also calculate the thread time for different statuses of the smart device, e.g., Idle, Active, and under attack.



\section{Static Analysis of Resource Usage Attack}
\label{SEC:SA}
The smart devices used in this experiment were infected with
\begin{figure}[ht]
	\includegraphics[width=\linewidth]{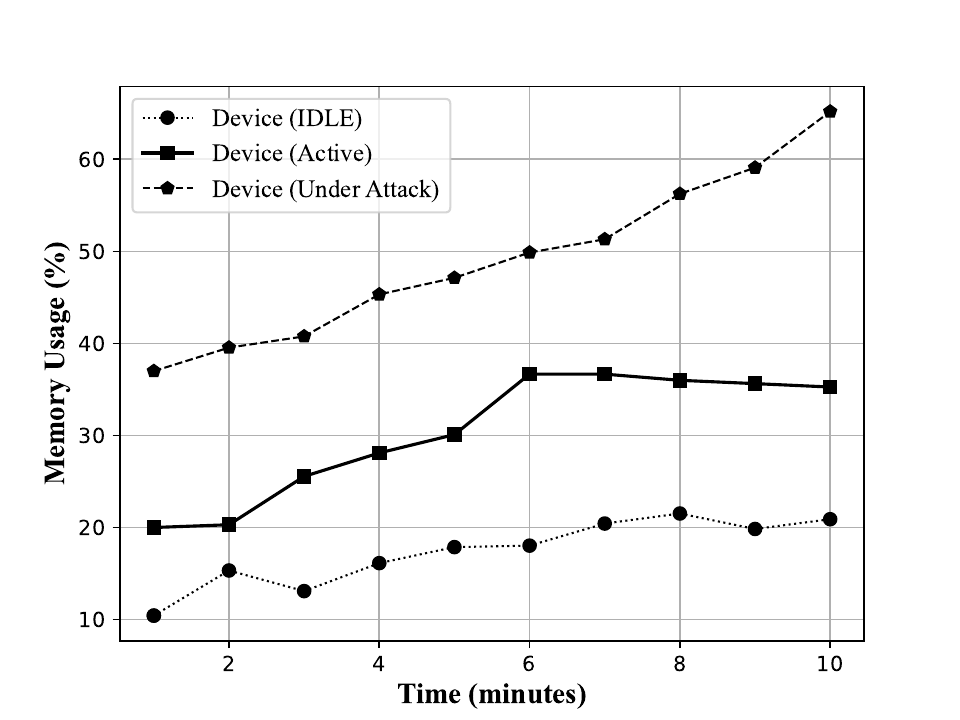}
	\centering
	\caption{Raspberry Pi (Memory Usage Before and After the attack).}
	\label{FIG:MUB}
\end{figure}
 malicious software used to carry out different malicious attacks on a target on an isolated network. During the experiments, the memory usage footprints of the smart devices were obtained under normal operating conditions, as well as when these smart devices carry out cyberattacks. Each memory usage footprint was obtained by taking measurements after $5$~s within $1$ minute when the smart device performs an attack and normal operation. A total of $10$ minutes of calculation measurement of memory usage footprints of both in the presence of attacks and normal functioning smart devices were built. 

During packet collection, the attacks are sent using the same Transmission Control Protocol (TCP) and User Datagram Protocol (UDP)
flood commands. Using the topology
as depicted by Figure~\ref{FIG:T2D}, the malicious TCP and UDP traffic are separately sent to the victim device, while all usage statistics are recorded on the victim device.
Each attack is simulated for a duration of $1$ minute, and all usage statistics are recorded for the same duration.
No attacks are sent during the first period ($10$ minutes), and all usage statistics are recorded and saved in the DB. And also, the same things applied once the second period started after sending malicious attacks.

\begin{figure}[ht]
	\includegraphics[width=\linewidth]{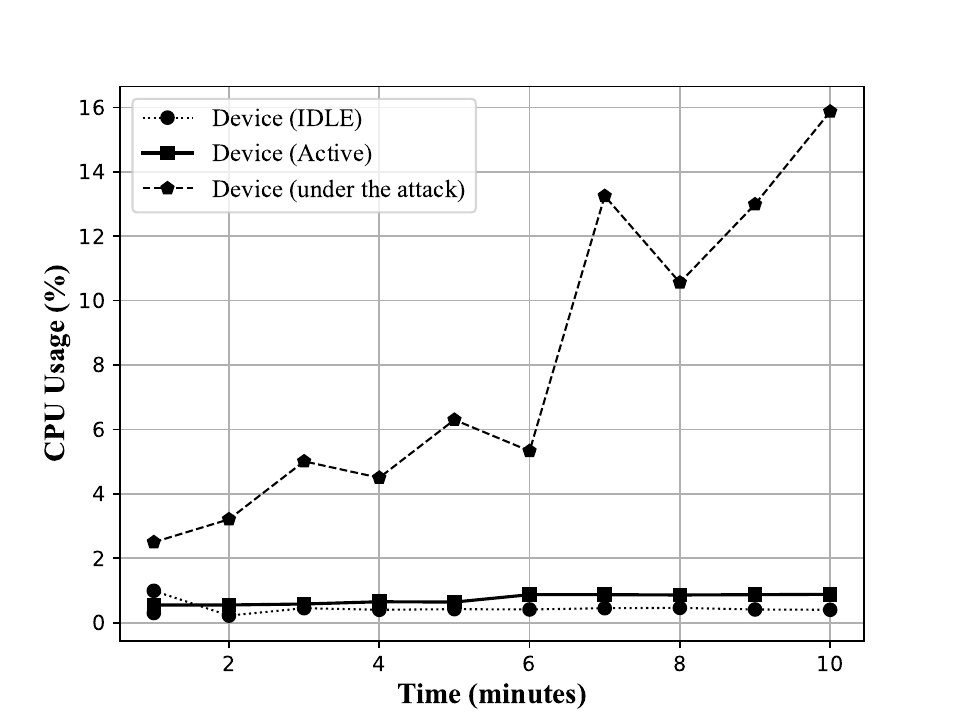}
	\centering
	\caption{Raspberry Pi (CPU Usage Before and After the attack).}
	\label{FIG:CPB}
\end{figure}

\begin{figure}[ht]
	\includegraphics[width=\linewidth]{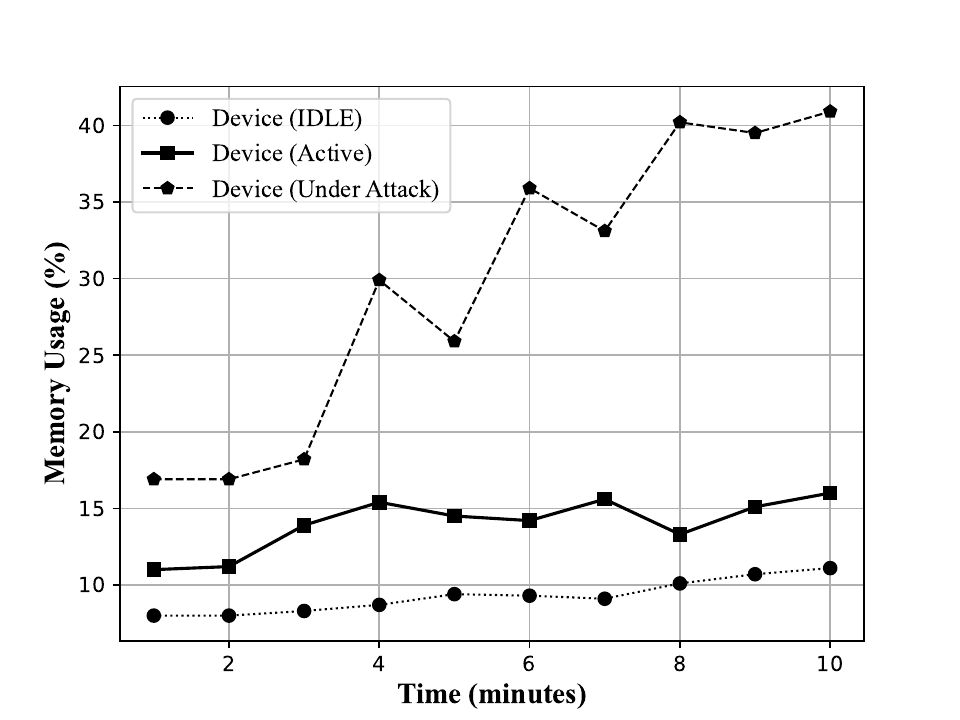}
	\centering
	\caption{Arduino (Memory Usage Before and After the attack).}
	\label{FIG:ARD}
\end{figure}

The result of this experiment shows the memory usage footprint when the device is Idle, Active, and under attack. Therefore, the normal usage of the memory of the Raspberry Pi device in the absence of the attack fluctuates between $10$\% to $36$\%. This percentage is divided between two different states of the smart device when it is Idle, the percentage is between $10$ and $20$\%, and when it is Active, the percentage is more than $25$\% but less than $37$\% as shown in Figure~\ref{FIG:MUB}. Moreover, the percentage of memory usage changed after sending the malicious attack, and the percentage changed to be more than $66$\% per minute. We also calculate the CPU usage of the smart devices to check the CPU status before and after attacking the memory of the smart devices. Figure~\ref{FIG:CPB} shows the normal CPU usage for Idle and Active statuses of the Raspberry Pi device. The normal CPU usage for Idle devices is between $0.55$\% to $0.88$\%. The memory usage of the Active smart devices is between $0.88$\% to $1.50$\%. At the same time, the CPU usage is more than $1.5$\% once we send the malicious attack to the smart devices.  
We also calculate the memory usage of another smart device (Arduino). The main purpose of using two different devices is to show how the algorithm works for different devices which implement different architectures. For printing the memory usage of the Arduino device, we used a specific library to get the free usage memory for different statuses of the smart device, e.g., Idle, Active, under attack. Therefore, the memory usage for the first status, as shown in Figure~\ref{FIG:ARD}, fluctuates between $8.1$\% to $11$\%, and for the Active status, it is between $11$\% to less than $16$\%. The memory usage percentage changes to more than $17$\% and less than $50$\% once we send a malicious attack to the smart device. 

The results and analyzes of this experiment assisted us in understanding the impact of the memory usage attack on smart devices and building a lightweight algorithm to protect these devices from such an attack. The following section describes the detection algorithm and presents some results.
\section{Threat Mitigation}
This paper introduces a detection mechanism and response to cyber-attacks on smart devices' memory usage. We also propose a lightweight algorithm to detect such memory changes inside smart devices by monitoring memory usage. Once the attack is detected, the algorithm will force the memory to stop listening to such an attack (e.g., stop reading and writing to memory). We also disconnect the victim devices from the Internet automatically. We implement this algorithm in the smart devices themselves. The presented mechanism records the response of the attack, and memory usage, for different states such as Idle, Active, under attack. The detection algorithm detects any breach in the memory usage of smart devices. 

\subsection{Proposed Algorithm}
\label{SEC:PA}
The attacker aims to consume more memory usage of the smart device, and the monitoring mode of the presented algorithm updates and registers all different cases of memory behaviors before and after the attack. We record the change on memory for every $3$ second for $1$ minute. According to the data obtained from the testbed, the attacker can change the memory usage within $67$\% of wrong values during $10$ minutes in total. 

\begin{algorithm}
\label{Algo:algo}
\caption{A Technique to detect Memory Usage Attack}
\begin{small}
\begin{algorithmic}[1]
    
    \State Input: $d,Diff, C1, T1, Alert$
    
    \State Output1: Normal($M1$) 
    \State Output2: Abnormal ($M2$) 
    \State Final Result: Output1 $or$ Output2
    \State $M1: Reading_{memory-usage}$
    \State $MEM(d) =f(mem(d),ATK,n)$
    \State $M2 : Reading_{memory-usage}$
    \State $Reading_{Threshold}= Max_{usage(MEM)}-Min_{usage(MEM)}$
    \State $Diff= M1 - M2$
    \If{$M2 = M1$} 
        \If{$Diff >Reading_{Threshold}$}
            \State $Reset T1$
            \If{$Alert == 'On'$}
                \State monitor memory
            \Else
                \State $C1=C1+1$
                \If{$C1 > Max_{(memory-usage)}$}
                    \State $Alert =='On'$
                    \State Attack detected
                    \State Detect the main source ($X$)
                    \State Stopped Reading/Writing on Memory from $(X)$ 
                    \State Disconnect the smart device ($d$) from the Internet
                \Else
                    \State Return back to monitor memory
                \EndIf
            \EndIf
        \Else
            \If{$C1 > 0$}
                \State $T1=T1+1$
                \If{$T1> Threshold_{T}$}
                    \State Reset Alert
                    \State Reset C1
                    \State Reset C1
                    \State Attack stopped
                \Else
                    \State return back to monitor memory
                    
                \EndIf
            \EndIf
        \EndIf
    \EndIf
    
\end{algorithmic}
\end{small}
\end{algorithm}
This Algorithm~\ref{Algo:algo} takes the recorded readings from the DB for each smart device in the IoE system. The variable $Diff$ stores the subtraction of previous (\emph{before sending such attack}) and current (\emph{after sending malicious attack}) smart devices reading. For instance, the maximum memory usage of the smart devices for Idle and Active smart devices are given in Figure~\ref{FIG:MUB}. 
The variable $Diff$ stores the subtraction of the previous and current memory usage readings. For instance, the maximum sudden memory usage change expected in the memory of the smart device is given by
subtracting the value of the maximum memory usage when the device is under attack minus the minimum memory usage when the device is Active and Idle before sending any attack.

\begin{equation}  \label{eq4}
    Reading_{Threshold}= Max_{usage(MEM)} - Min_{usage(MEM)}
\end{equation}

When the variable $Diff$ exceeds the expected value, the variable $T1$ \emph{"Timer"} is reset, and we verify whether the alert message has been sent to the administration. If not, we increase the $counter1$ variable, which records the number of times the difference between previous and current memory usage reading exceeds the maximum allowed value. When the $counter1$ variable is greater than the maximum allowed value, it sends the alert message indicating that the space of memory usage addressed to that smart device is under memory usage attack. At this stage, we stop any reading/writing operations to and from memory and disconnect the smart device from the Internet, as all victim device's IPs are stored in the black-list of our DB.

Through experimentation, we consider the scenario when the attacker stops the attack. When the $Diff$ value is less than $Reading_{threshold}$ value, we compare whether the variable counter is greater than $zero$, then we increase $T1$. We can assume the attack stops if the variable is greater than the $Time_{Threshold}$ variable. Finally, we reset the alert: $counter1$ and $T1$ variables.

After detecting the memory usage attack of such a device $(d)$, we put all the victim devices on a black-list. Then, once the attack is detected on such a device, we first stop any operation on the memory, e.g., read and write on memory. We disconnect the Internet connection of the smart device $(d)$ to prevent any further attack on the smart device's memory usage. The next section presents different results regarding detecting memory usage attacks.

Therefore, the mitigation is summarized in the following steps:
\begin{enumerate}
    \item add the victim smart devices' IP to a black-list;
    \item stop any reading/writing to the smart device;
    \item disconnect the smart device from the Internet.
\end{enumerate}

\section{Experimentation and Discussion}
\label{SEC:EXP}

\subsection{Results}
\label{SUB:RESults}
\begin{figure}[!ht]
	\includegraphics[width=\linewidth]{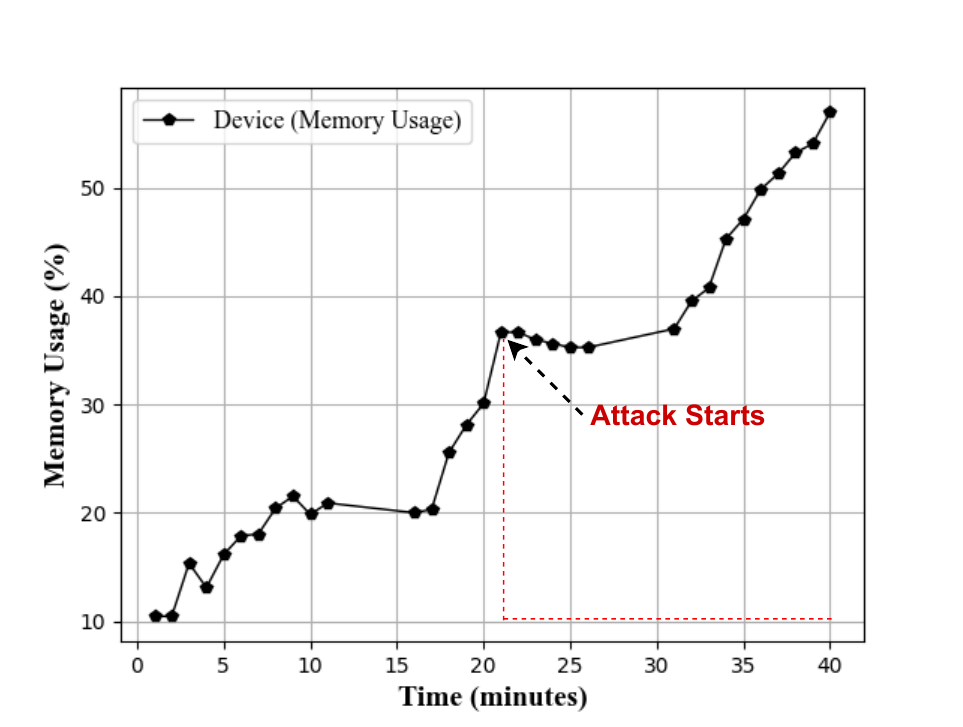}
	\centering
	\caption{Raspberry Pi (Detecting the memory usage Attack.)}
	\label{FIG:D1}
\end{figure}

We ran malicious attacks on the smart device to check the memory usage before and after the attack. Figure~\ref{FIG:D1} shows the mechanisms of our algorithm to fetch the attack once it is started. The monitoring mode of the memory usage sends memory usage readings to the algorithm, and inside the algorithm, there is a statistics comparison between normal and abnormal cases. As described in Section~\ref{SEC:TS}, we first check the behavior of the smart devices once there is an attack, and we register all different cases for memory usage, e.g., Idle, Active, under attack. The main purpose of this analysis is to study the attack first and then to build a mitigation mechanism to detect memory attacks.  

\begin{figure}[ht]
	\includegraphics[width=\linewidth]{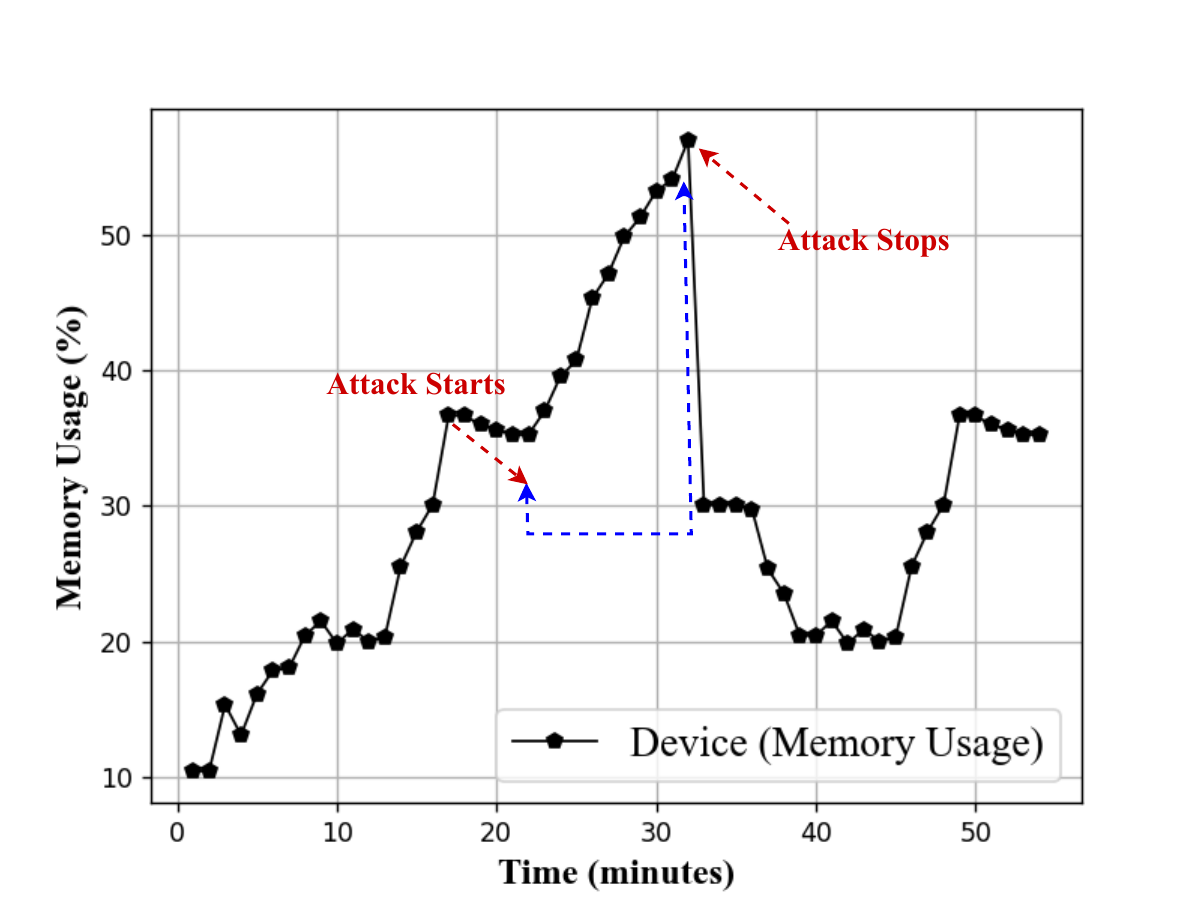}
	\centering
	\caption{Raspberry Pi (Detecting the memory usage once the attack starts and when it stops).}
	\label{FIG:D2}
\end{figure}

Figure~\ref{FIG:D1} and~\ref{FIG:D2} shows the presented results of detecting the attack once it starts; we can notice that the attack starts when the memory usage is greater than $37$\%, and the $Diff$ variable is greater than the expected memory usage value. At this stage, the smart device $d$ is passed through different operations, e.g., stop reading/writing on $d$, disconnect $d$ from the Internet to stop any further attack, and send an alert to the administration about the status of the smart device. 

\begin{figure}[ht]
	\includegraphics[width=\linewidth]{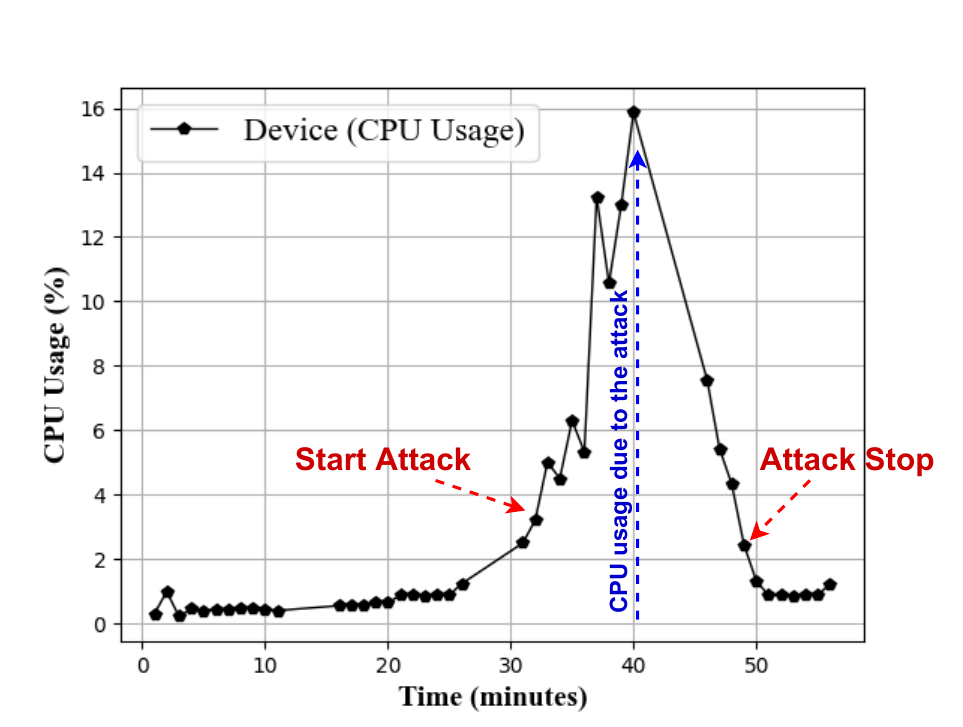}
	\centering
	\caption{CPU Usage during the attack when it started and stopped (Raspberry Pi).}
	\label{FIG:CPUD1}
\end{figure}

The detection algorithm also notified the administration once the attack stopped. This stage will help with further operations. 
Through this experiment, we also studied the behavior of the CPU usage of the Raspberry Pi device under the same attack. Figure~\ref {FIG:CPUD1} shows the behavior of the CPU usage before and after attacking the memory of the smart devices. We also applied the same detection algorithm to study the behavior of the mitigation algorithm on the CPU and whether this algorithm detects the attack or not. 

\begin{figure}[ht]
	\includegraphics[width=\linewidth]{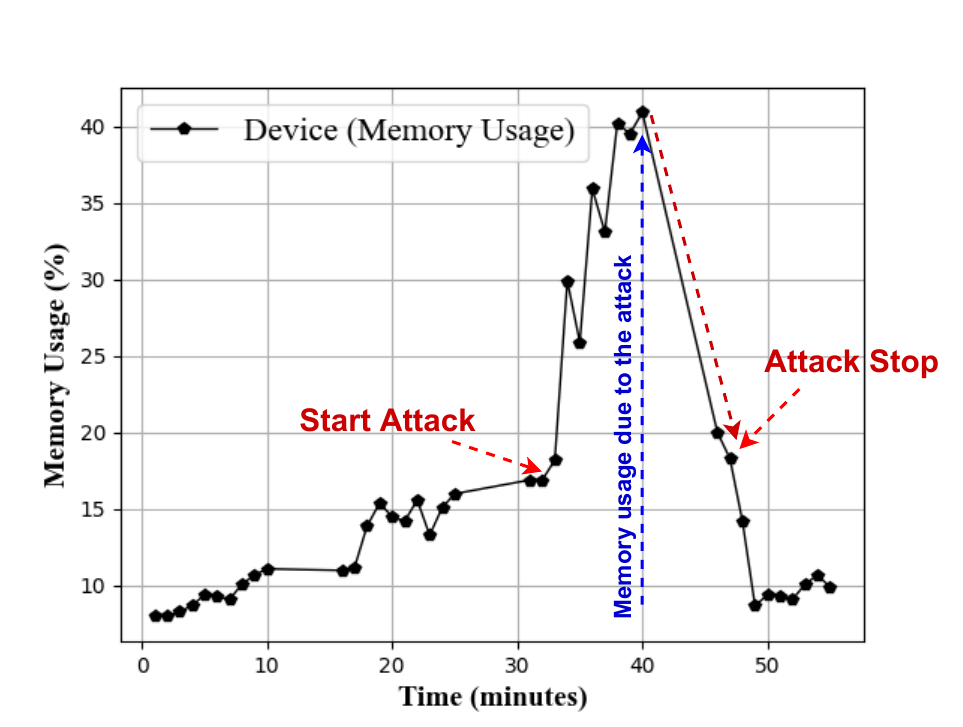}
	\centering
	\caption{Arduino (Detecting the memory usage once the attack starts and when it stops).}
	\label{FIG:ArdD2}
\end{figure}

The same calculation is applied to the Arduino, and the detection algorithm records different variables about the attack once it is started and stopped. Figure~\ref{FIG:ArdD2} shows the recorded results of detecting the attack. We can notice that the attack started when the memory usage percentage increased to be more than $16$\%, and for detecting the attack when it is stopped, once the memory usage percentage decreased to be less than $20$\%. Once the system detects that the attack is stopped on the smart device $d$, the system might disconnect the smart device from the Internet, or the actual attack is stopped from the main source. 

The algorithm also stores all victim devices' IPs in the black-list, so when there is an attack on the smart device, we disconnect the smart device to prevent any further attack. We also prevent further access to the database until the administration team solves the issue. 
\section{Conclusion and Future work}
\label{ConFU}

The Internet of Everything is the beginning of a new era of technology in Internet-based smart communication and connecting smart devices. The security of IoE pillars is important as some suffer from resource-constraints problems. This paper proposed an approach that can detect and classify memory usage attacks using memory-based features extracted from the memory usage of the smart device. The approach represents a mitigation method to detect the attack once it appears in the memory usage of the smart devices. First, we monitor memory usage by using a specific tool in \emph{Python} script and \emph{C} language to fetch different data about memory usage. Then, all the fetched data is stored in the DB for further calculation. Second, we studied the attack behavior and registered the memory usage readings before and after the attack. In this work, we conduct static and dynamic analysis of the memory usage attack. In particular, we have conducted all the
experiments in an isolated and cost-efficient experimental
setup. It is observed that malicious attacks, e.g., flooding attacks have a significant impact on the resources of the IoE smart devices. When an IoE edge device is flooded with malicious attacks, there are significant increases in CPU and memory usage. This analysis helps in building the detection algorithm. The detection method relies on monitoring the memory usage to compare different variables of the memory reading. It is also able to detect the attack on time once it happens. Moreover, it can detect if the intruder stops the attack or not. We also build an alert message inside the algorithm to send different notifications to the administration once the attack is detected. Moreover, all victim devices are disconnected from the Internet, and all read/write operations to and from memory are also stopped. In the future, we will focus on detecting the main sources of memory usage attacks in the IoE environment.

\bibliographystyle{splncs04}
\bibliography{bibliography}
\end{document}